\documentclass[acmsmall,screen,nonacm]{acmart}

\usepackage[normalem]{ulem}
\usepackage{xurl}
\usepackage{listings}
\usepackage{algorithm}
\usepackage{algpseudocode}
\usepackage{styles/c_custom_style}
\usepackage{xspace}
\usepackage{microtype}
\usepackage{balance}
\usepackage{amsmath}

\algnewcommand\algorithmiccontinue{\textbf{continue}}
\algnewcommand\algorithmicswitch{\textbf{switch}}
\algnewcommand\algorithmiccase{\textbf{case}}
\algnewcommand\algorithmicassert{\texttt{assert}}
\algnewcommand\Assert[1]{\State \algorithmicassert(#1)}%
\algdef{SE}[SWITCH]{Switch}{EndSwitch}[1]{\algorithmicswitch\ #1\ \algorithmicdo}{\algorithmicend\ \algorithmicswitch}%
\algdef{SE}[CASE]{Case}{EndCase}[1]{\algorithmiccase\ #1}{\algorithmicend\ \algorithmiccase}%
\algtext*{EndSwitch}%
\algtext*{EndCase}%

\newtheorem{definition}{Definition}

\newcommand*\ttvar[1]{\texttt{\expandafter\dottvar\detokenize{#1}\relax}}
\newcommand*\dottvar[1]{\ifx\relax#1\else
  \expandafter\ifx\string_#1\string_\allowbreak\else#1\fi
  \expandafter\dottvar\fi}

\newcommand{\shortname}{\textsc{PDSG}\xspace}

\newcommand\dlimpl{\operatorname{:-}}

\begin{document}

\title{Combined Static Analysis and Machine Learning Prediction for Application Debloating}

\author{Chris Porter}
\email{porter@ibm.com}
\affiliation{
  \institution{IBM Research}
  \city{Yorktown Heights}
  \state{NY}
  \country{USA}
}

\author{Sharjeel Khan}
\email{smkhan@gatech.edu}
\affiliation{
  \institution{Georgia Institute of Technology}
  \city{Atlanta}
  \state{Georgia}
  \country{USA}
}

\author{Kangqi Ni}
\email{kni3@gatech.edu}
\affiliation{
  \institution{Georgia Institute of Technology}
  \city{Atlanta}
  \state{Georgia}
  \country{USA}
}

\author{Santosh Pande}
\email{santosh.pande@cc.gatech.edu}
\affiliation{
  \institution{Georgia Institute of Technology}
  \city{Atlanta}
  \state{Georgia}
  \country{USA}
}

\begin{abstract}

Software debloating can effectively thwart certain code reuse attacks by
reducing attack surfaces to break gadget chains. Approaches based on static
analysis enable a reduced set of functions reachable at a callsite for
execution by leveraging static properties of the callgraph. This achieves low
runtime overhead, but the function set is conservatively computed, negatively
affecting reduction. In contrast, approaches based on machine learning (ML)
have much better precision and can sharply reduce function sets, leading to
significant improvement in attack surface. Nevertheless, mispredictions occur
in ML-based approaches. These cause overheads, and worse, there is no clear way
to distinguish between mispredictions and actual attacks.

In this work, we contend that a software debloating approach that incorporates
ML-based predictions at runtime is realistic in a whole application setting,
and that it can achieve significant attack surface reductions beyond the state
of the art. We develop a framework, Predictive Debloat with Static Guarantees
(\shortname). \shortname is fully sound and works on application source code. At runtime it
predicts the dynamic callee set emanating from a callsite, and to resolve
mispredictions, it employs a lightweight audit based on static invariants of
call chains. We deduce the invariants offline and assert that they hold at
runtime when there is a misprediction. To the best of our knowledge, it
achieves the highest gadget reductions among similar techniques on SPEC CPU
2017, reducing 82.5\% of the total gadgets on average. It triggers misprediction
checks on only 3.8\% of the total predictions invoked at runtime, and it
leverages Datalog to verify dynamic call sequences conform to the static call
relations. It has an overhead of 8.9\%, which makes the scheme attractive for
practical deployments.

\end{abstract}

\date{}
\maketitle

\section{Introduction} \label{sec:introduction}

Modern software contains a substantial amount of unused code
\cite{piece-wise,green-bloat,debloat-art,debloat-study},
and this unneeded code can be leveraged in code reuse attacks
\cite{Nergal,expressiveness_retlibc,rop,jop,pcop}.
In particular, some parts of modern software deal
with very rare corner cases or highly rare exceptional
conditions which are not regularly exercised, nor
updated and maintained. These parts of software can contain
exploits, but recently software debloating has made progress
in adding defenses for such cases.
The idea is to remove as much of this excess code
as possible in order to reduce the program's attack surface.

Table \ref{table:tool-comparison} shows \% reductions in gadgets
for several debloating frameworks since 2018.
As shown in ~\cite{decker},
attackers typically use a memory vulnerability along with a
chain of gadgets to construct a code reuse attack. 
Section \ref{sec:background} gives details of gadget-based attacks, but for now,
gadgets can be understood as the basic building blocks for
advanced code reuse attacks; attackers stitch gadgets into
a dynamic chain that transfers control and data-flow (illegally)
from one code location to another, which can culminate
into complete control over program execution
(through a mechanism such as an $execve$ call).
The ``Debloat'' column in Table \ref{table:tool-comparison}
gives a rough comparison of 
the number of gadgets removed by a given framework.
Although the frameworks are evaluated on different benchmarks,
have differences in terms of their targeted artifacts (applications vs. libraries),
or may be static or dynamic techniques, 
one of the most striking aspects of
Table \ref{table:tool-comparison} is BlankIt's \cite{blankit}
high debloat percentage.
It reduces the number of gadgets that are active at runtime by
97.8\%. Though direct comparisons are strained, the next closest gadget
reduction result is for Decker, which is $\sim$25\% worse.
It may be noted that BlankIt uses ML-based prediction at runtime,
whereas Decker is based on static analysis.

\begin{table}[ht!]
 \centering
  \caption{Recent debloating frameworks and how much ``debloat'' they perform.}
 \begin{tabular}{|l|l|l|l|}
  \hline
  \textbf{Year} & \bf Tool & \bf Debloat &  \bf Notes \\
  \hline
  \hline
    2018	&	Trimmer	&	20.0\%	&	Total gadget reduction \\
    2018	&	Chisel	&	66.2\%	&	ROP gadget reduction \\
    2018	&	Piece-wise	&	72.9\%	&	Total gadget reduction \\
    2019	&	Razor	&	68.2\%	&	Code reduction \\
    2020	&	BlankIt	&	97.8\%	&	ROP gadget reduction \\ 
    2022	&	LMCAS	&	55.9\%	&	ROP gadget reduction \\ 
    2023	&	Decker	&	73.2\%	&	Total gadget reduction \\
  \hline
  \end{tabular}
  \label{table:tool-comparison}
\end{table}

\subsection{ML-based Prediction for Debloating: Pros and Cons}
Upon closer analysis between BlankIt and Decker, we come to the conclusion
that the key factor for higher gadget reduction in BlankIt over Decker
is that BlankIt uses ML-based prediction, which is very precise. 

BlankIt's ML-based predictions are input-sensitive and
can leverage calling context to forecast
the next set of calls in a dynamic call chain.
Such predictions can overcome
the unanalyzable nature of static callgraphs and control flow graphs.
Using ML in a debloating framework carries with it
one key challenge, however, namely {\bf how to resolve mispredictions}.
BlankIt, for example, suffers from this problem of false alarms.
When a library call leads to a misprediction, BlankIt attempts
to resolve it in a parallel
audit thread that reruns the call within
Memcheck \cite{Nethercote:2007:VFH:1250734.1250746}.
Though serviceable for library code, this technique would
not be tenable for predicting function calls on a whole program basis in
application code, which is what Decker deals with.
Library calls are relatively infrequent compared to
user-defined functions, so mispredictions are likely to
occur with higher frequency when predicting and debloating for
application code on a whole program basis.
In other words, the use of ML-based prediction and debloating 
would lead to a situation with constant interruptions (due to mispredictions)
and audit threads for distinguishing
between attacks vs. mispredictions.
Therefore, new techniques are
needed that are lightweight and can validate predictions
without such problems.
\textit{In short, it remains an open problem how to effectively
verify that mispredicted execution paths still conform to a valid
control flow}.

\subsection{Limitations of Static Callgraph-based Prediction for Debloating}
Although Decker provides a significant attack surface reduction,
being based on static analysis of callgraph reachability comes with limitations. 
We draw on a simple code example to highlight how static analysis-based
callgraph predictions can fail to protect the program,
and how a technique with ML-based predictions provides a higher
level of defense. Listing \ref{code:motivatingex}
shows pseudocode that contains a common programming paradigm,
namely error handling for extremely rare corner cases.
In this example, the function $service$ contains a while-loop
for $do\_work$. On each iteration, the loop checks if
error $e$ has occurred;
if so, $handle\_err$ is called to process the state;
then execution continues in $do\_work$ for the current iteration.

The intuition for this simple but realistic code is that the error handler
does not always need to be mapped active, because it is unlikely
to be called.
Based on static reachability analysis triggered at the loop header,
Decker \cite{decker} fails to handle this case, though,
because its static analysis is conservative.
In Decker, this code is handled by first statically
analyzing the loop and identifying all interprocedurally
reachable functions within it ($do\_work$, $aux1$,
$aux2$, $handle\_err$, and $attacker\_target\_func$). Then
at runtime, Decker enables this entire set of functions at
the loop preheader and disables these functions at the loop exits.
A technique such as BlankIt \cite{blankit} cannot be used here
because it is only able to handle library code, as explained above.
If, however, one can use a model to predict which functions
will be called within this loop, it is reasonable to expect
that $handle\_err$ will not be regularly predicted
when the model is trained on representative profiling data for this case.

We list three ways in which
an attacker may try to enter and exploit $attacker\_target\_func$,
and describe how a predictive model with statically enlightened
checks can defend against them:
\begin{enumerate}
  \item The target function may contain a gadget that
  the attacker needs to use in their gadget chain.
  Under the most common executions, however, a model
  will predict that $attacker\_target\_func$ will not be executed,
  so this function will be disabled. Jumping into this code would
  therefore lead to a fault and a failed attack, unlike in Decker.
  \item The attacker may attempt to redirect control flow
  to the entrance of $attacker\_target\_func$. Though this is
  a normal function entry point (unlike case 1), it will
  still lead to a fault and crash whenever the model has
  not predicted the error handler to be invoked.
  (Section \ref{sec:rectification} will elaborate on this point.)
\item The attacker may attempt to redirect control flow to
  the entrance of $handle\_err$. As we will show, all entry points
  into subsections of the callgraph that depend on the model
  predictions must be gated by a fallback mechanism. This
  is to handle mispredictions under no-attack scenarios.
  However, the static analysis techniques we introduce later
  can catch such cases. In particular, only
  a valid static sequence of calls will be allowed to pass
  through this ``gate.'' In our example, the attack will
  only be allowed to jump to $handle\_err$ from $aux2$
  (as $handle\_err$ can only follow $aux2$ in a valid
  call function sequence). Attempting to jump from $service$,
  $do\_work$, or $aux1$ will fail the runtime
  check at the entrance to $handle\_err$ and result in a crash.
  (Sections \ref{sec:rectification} and \ref{sec:path-checking}
  will make these points clearer.)
\end{enumerate}

\begin{lstlisting}[
  xleftmargin=.05\textwidth,
  xrightmargin=.05\textwidth,
  caption={Pseudocode illustrating common programming paradigms with a do-work loop and error handling.},
  label=code:motivatingex,
  language=python,
]
def service():
  while p:
    e = do_work()
    if e:
      handle_err()
def do_work():
  aux1()
  return aux2()
def handle_err():
  attacker_target_func()
\end{lstlisting}

\subsection{Contributions}
With the above motivations, we present a new ML-based predictive
method backed by performant audits that use static invariants. 
In summary, this work makes the following contributions:
\begin{enumerate}
    \item To the best of our knowledge, it is the first prediction-based approach
    for whole-application debloating that is performant and sound.
    \item It is the first debloating technique to leverage static analysis
    techniques that guarantee certain program properties are met whenever mispredictions
    due to ML occur.
    \item It includes an empirical evaluation that shows the technique improves
    attack surface reduction beyond the state of the art and with overheads that
    are in line with prior art.
\end{enumerate}

The remaining sections are laid out as follows.
Section \ref{sec:background} provides background.
Section \ref{sec:overview} gives an overview of the solution.
Section \ref{sec:framework} goes into technical details of the framework.
Section \ref{sec:evaluation} provides the results from our empirical evaluation.
Section \ref{sec:related-work} provides a survey of the related work.
Section \ref{sec:conclusion} concludes.
\section{Background}\label{sec:background}

We further motivate our work by delving into more details 
of the research artifacts in Table \ref{table:tool-comparison}.
Additional related work is in Section \ref{sec:related-work}.
We also provide a brief outline of gadget-based attacks
and the threat model for this work at the end of this section.

\subsection{State-of-the-art Debloating}

Designing a debloating tool involves tradeoffs along different dimensions.
A debloating technique can have several attributes 
including  soundness,
ability to offer full feature support, whether it
requires user input or manual intervention, whether it
depends on a runtime component,
etc. Regardless of the technique used, the most important
criteria for a tool in terms of real-world utility are
soundness under a normal (no-attack) scenario and
the ability to stop an attack with acceptable runtime overhead.
The soundness criterion is operationally defined as 
the ability of the debloated program to be robust and
not crash under no-attack inputs, and to produce the same
results as its non-debloated counterpart under such inputs.
By full feature support, we mean the debloated program is
not ``specialized'' to a subset of features from the
original program; it must provide support for all the desired
(non-debloated) features.
In this work, we maintain that soundness and full feature
support are critical for a generally adoptable solution.
Our goal is to show how a mix of static analysis techniques
and machine learning can improve debloating reductions
without sacrificing either of these qualities.
In addition, we also show how the attack-stopping ability
is superior to Decker and runtime overhead lower than BlankIt.

There are only three tools from Table \ref{table:tool-comparison}
that are sound and that produce programs with full feature support:
Piece-wise \cite{piece-wise}, BlankIt \cite{blankit}, and Decker \cite{decker}.
Despite this commonality, these tools differ substantially.
For example, Piece-wise and BlankIt
debloat libraries; Decker debloats applications.
BlankIt is the only tool of this group with a ML-based
prediction mechanism;
it is also the only one that works at the binary level.
Piece-wise does not require any runtime support;
BlankIt and Decker do.

Piece-wise and Decker both show that to guarantee
soundness, the framework is forced to reckon
with the limitations of static analysis. In fact, Trimmer \cite{trimmer}
and LMCAS \cite{lmcas}, despite being specialization techniques,
face this same issue in order to guarantee soundness.
Razor \cite{razor} is also a specialization technique;
it runs the program on representative inputs and then uses
heuristics to include related control flow; unlike the above
techniques, however, it is unsound.


There are two frameworks in Table \ref{table:tool-comparison} that
rely on ML-based prediction: Chisel \cite{chisel} and BlankIt.
In Chisel, the model makes offline decisions on where to cut parts of the
program, which may lead to unsound transformations;
once the program is finalized, there
is no support to handle missing functionality at runtime.
In contrast, BlankIt's runtime allows it to access debloated code even on misprediction;
it depends on
Memcheck \cite{Nethercote:2007:VFH:1250734.1250746,DBLP:conf/usenix/SewardN05}
to catch memory errors in this case.
Though the interfaces and frequency of third-party library
calls may enable this kind of mitigation approach in BlankIt,
it is unlikely to be useful for whole program debloating which is the focus of this work.
The authors' results show that to reduce slowdowns on the critical
path, a parallel audit thread is required, but even in two
of their evaluated benchmarks, more than one helper thread is needed
to handle the audit overhead and frequency of mispredictions.


\subsection{Gadgets and Code Reuse Attacks}
Gadgets are a continuous sequence of instructions
that perform some useful computation for an attacker. It could be a primitive
action (such as addition) or a more complex one (such as pushing
onto a stack and advancing an index). It may also have some side
effect (whether wanted or unwanted) such as polluting a register that
is unrelated to the desired computation. In any case, a gadget in
isolation is rarely useful, so attackers combine them to form gadget
chains. The gadget chain carries out some malicious action,
such as spawning a shell or changing the permissions of a
system page. Since such attacks are carried out by reusing the
application code (in terms of gadget chains), they are called code reuse attacks. 

Chaining gadgets requires some linking mechanism to go from one gadget
to the next during an attack. For example, return-oriented programming
(ROP) \cite{rop} uses return instructions, jump-oriented programming (JOP)
\cite{jop,jop2} uses jump, and call-oriented programming (COP) \cite{pcop} uses call.
There are nuances to each of them. Jump-oriented programming commonly
uses a dispatcher function, for example. There are also flavors
within each of these types, such as JOP attacks that use two
dispatcher functions \cite{jop-rocket}. In fact, ROP, JOP, and COP
can even be intermixed in a gadget chain.


Lastly, it should be noted that using gadget reductions
as a debloating measurement comes with limitations.
Works such as \cite{islessreallymore} show the inherent weaknesses of reduction
metrics and present a tool for analyzing gadget classes and quality.
By considering what various gadgets do or compute, the kind and extent
of their side effects, how critical they are, how rare they are, etc.,
one can form a more complete picture of the effectiveness of a
debloating technique. In short, measuring only gadget quantity
oversimplifies all of this. Furthermore, as mentioned,
gadgets must be strung together into a useful chain. Reducing
the absolute gadget count in a program may not matter if a critical
chain still exists and can be exploited.

These limitations are an open problem within the debloating for security
community. Automatically constructing gadget chains is challenging,
though some tools exist \cite{ropgadget, ropper}. In \cite{decker},
the authors show that relatively high gadget reductions are capable of
breaking a well-known shell-spawning chain on common benchmarks.
This promising result suggests that high gadget-reduction counts
can thwart certain known chains.

We limit our scope of work solely to the problem of how to
improve results using known gadget metrics. We focus
on the problem of eliminating more code surface. Therefore,
we use similar metrics from prior art, including gadget classification,
count, and quality, but we acknowledge and treat the metrics problem
itself as important future work.


\subsection{Threat Model}

We adopt a threat model similar to previous debloating
works \cite{chisel, razor, decker, blankit, lmcas}.
The underlying hardware, software stack, and toolchain are
trusted. The threat is an attacker carrying out a code reuse
attack on the application itself. We do not
make any assumptions about the initial vulnerability or exploit
for triggering the code reuse attack. We are
focused only on reducing the attack surface of the program such
that any exploit will have substantially less code and gadgets available
for reuse. We assume the runtime is protected as in similar
works that use existing methods, e.g. hardware segmentation
\cite{blankit, decker, cpi, getting-point}).
Similar to works like \cite{decker}, the goal is attack surface
reduction and is not about guaranteeing the integrity of indirect
control flow (as this is already covered in numerous CFI \cite{cfi}
works).

\section{Overview} \label{sec:overview}

The \shortname framework is a mix of compiler and runtime components
that work in tandem to reduce the attack surface of application code.
The static component consists of LLVM compiler passes and
Datalog analysis. The dynamic component is implemented as a user-level
runtime library to support logging (when training a predictive model)
and code protection during an actual run of the application.
The general approach of the framework is to build a program that
at runtime marks forthcoming functions as executable
just before use and marks them read-only afterward.
In this manner, only the functions that are enabled executable
at any particular moment during execution will form part
of the attack surface; and if an attacker tries to jump to any other
region of code, that code will be read-only and cause the program
to crash.
A predictive model helps to narrow this attack
surface by guiding this step. Only application functions that the
model predicts will be invoked are marked executable.
Because of mispredictions, however, additional instrumentation
and runtime support is needed to separate mispredictions from actual
attacks (more details in Section \ref{sec:framework}).

\shortname has several properties in common with similar prior
works (e.g. \cite{decker}).
When marking functions as executable or read-only, it works at the
granularity of system code pages.
It can reorder and align functions at build-time that are typically
called together.
It also resolves function pointers at runtime so that
attack surface does not explode due to conservative pointer analysis.

To use \shortname, an application goes through two stages: profiling and release.
In both stages, a user builds their application with \shortname's
compiler pass and then runs their application with \shortname's
runtime support. In the profiling stage, there is also a model-training
step, and that model is fed to the release stage.

\begin{figure}[ht]
    \centering
    \includegraphics[scale=0.4]{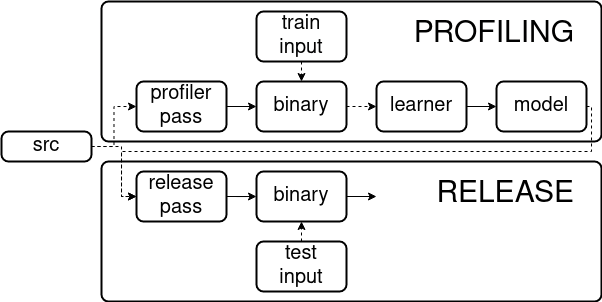}
    \caption{High-level flow diagram of \shortname's profiling and release stages. Dashed arrows indicate ``feeds into;'' solid arrows indicate ``creates.''}
    \label{figure:artd-high-level-flow-diagram}
\end{figure}

Figure \ref{figure:artd-high-level-flow-diagram} depicts a high-level
flow diagram of the framework.
The profiling stage is shown in the top half of the figure.
It is responsible for instrumenting the application
source code (using \shortname's custom LLVM profiler pass),
producing logs for model training
(by running the profiling binary on training input), and training the
model for the release stage (by using the logs generated during the binary's
training runs and passing this through a learner).
The release stage is shown in the bottom half of
Figure \ref{figure:artd-high-level-flow-diagram}. It is responsible
for instrumenting the application source code
(using \shortname's custom LLVM release pass and the model generated
during profiling), and running the binary on test input (i.e. real input
for exercising the model and debloating functionality).

\subsection{Release Stage Runtime Example}
\begin{figure}[ht]
    \centering
    \includegraphics[scale=0.44]{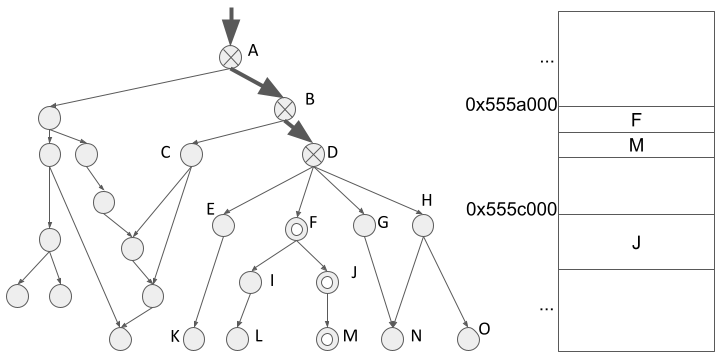}
    \caption{On the left, a callgraph with execution flow (X nodes)
    and prediction (donut nodes);
    on the right, a corresponding memory layout of the predicted functions.}
    \label{figure:predict-example}
\end{figure}

Listing \ref{code:predict-instrument} and 
Figure \ref{figure:predict-example}
illustrate the prediction mechanism at a high level.
In the listing, pseudocode for function $D$ is shown;
and in the figure, assume that execution has flowed
from function $A$ to $B$ to $D$, at which point
a prediction mechanism is triggered.
Notice that $D$ has a call to \texttt{predict} just above the while loop.
This \texttt{predict} call is
instrumented by \shortname's LLVM release pass. During the application's
execution, \texttt{predict} invokes \shortname's runtime system,
which queries the model for a set of forthcoming functions.
In this example, the runtime predicts $F$, $J$, and $M$.
The runtime marks these functions as RX before returning
from \texttt{predict} and entering the loop. The right side of
Figure \ref{figure:predict-example} makes it explicit that these functions
occupy some pages in system memory (not necessarily continuous or in order),
and it is only these upcoming pages of the memory that will be enabled
RX by this particular prediction.
In this manner, \shortname's runtime system is superior to that of Decker.
In Decker, the entire, statically reachable set of functions within
the while loop would be enabled ($\{E, K, F, I, L, J, M\}$),
whereas in \shortname, only \textit{predicted} functions would
be enabled ($\{F, J, M\}$ in this example), which can significantly improve
the precision and attack surface.

\begin{lstlisting}[
  xleftmargin=.05\textwidth,
  xrightmargin=.05\textwidth,
  caption={Example pseudocode for function $D$ to help demonstrate
  the release stage's instrumentation and prediction.},
  label=code:predict-instrument,
  language=C,
]
D(...) {
    ...
    predict(...)
    while(...)
        if(...)
            F(...)
        else
            E(...)
}
\end{lstlisting}

\subsection{Challenges}


The prediction techniques and runtime checks described above
raise several new challenges that must be solved for application code:
\begin{itemize}
    \item \textbf{Prediction}:
    It is unknown what kind of model is effective for this problem,
    what the features should be, and where the release pass should instrument
    the \texttt{predict} calls.
    \item \textbf{Rectification}:
    On misprediction, the program will
    attempt to call a function that is not yet mapped executable.
    \shortname must identify these ``rectification points'' in the program
    and map the necessary functions to avoid crashes. It is unclear which
    functions should be mapped at these points; how functions in the .text
    section should be laid out to accommodate this; and
    where the release pass should instrument these rectification points.
    \item \textbf{Path checking}: Finally, and most importantly,
    one must to distinguish between mispredictions and actual attacks.
    This raises several questions, including:
    Which properties of the program can and should be checked?
    How does one construct such a static model of
    invariants with reasonable compile time and that is also
    fast for runtime-checking?
\end{itemize}

\section{Technical Components} \label{sec:framework}
\shortname has three technical components, which address the three challenges
described at the end of Section \ref{sec:overview}:
prediction, rectification, and path checking.

\subsection{Prediction} \label{sec:prediction}
We consider a naive prediction scheme, followed
by a more robust approach, and then discuss
the model itself.

\subsubsection{Naive Prediction Scheme}
One could implement a method to predict callgraph behavior
without taking any of the callgraph structure into account.
We describe such an approach, because its shortcomings motivate the need
to develop a more sophisticated technique.

The idea of our naive approach is to treat the function calls
of an executing program as a sequence (ignoring program structure
entirely). That is, given a sequence of the last $X$ previously
executed functions,
the goal is to train a model that will predict the next $Y$ functions in the
future. Sequence-based prediction is a well-studied problem in
machine learning, and we choose one of its most
popular, basic solutions to model this problem,
long short-term memory (LSTM) \cite{lstm}. Using offline traces, we first study
the efficacy of such an approach.

We run SPEC CPU 2006 benchmarks with training input and with
function tracing enabled.
This outputs an ordered sequence of the executed functions.
We feed this as training data to a Keras implementation
of an LSTM, and this produces a trained model.
We then rerun the benchmark suite using its reference inputs,
and we use its output trace to test the model.
On average, the prediction accuracy is 91.1\%.

There are several critical shortcomings of these results. The accuracy
may be sufficient for the problem context ($\sim$10\% misprediction rate
could still be handled, potentially). To achieve these
results, however, the scheme takes several shortcuts.
The traces are very large (10s to 100s of GBs) -- so large
that it is unrealistic to expect users to capture them just
to compile moderately large applications. One engineering technique
could be to pipe these logs on-the-fly into a learner, which complicates
design but is feasible. Instead, we choose a simpler technique for this
naive scheme, which is to increase
the inlining threshold substantially (which reduces function calls
and therefore the log sizes). Note that this increases compilation
time and modifies the application beyond what the original programmer
may expect (i.e. is probably not tenable in a real-world scenario).
Nevertheless, this approach allows us to train a model for
most of the benchmarks in the suite.
Unfortunately, the logs are still too large for several benchmarks,
even on SPEC CPU 2006's smallest input. In these cases, we must
short-circuit during tracing and ignore certain callsites that
are invoked too frequently. This is an unrealistic strategy for
real-world applications, and it also negatively impacts prediction
accuracy.

Another serious shortcoming is how to reasonably handle the
mispredictions. Because of the sequential nature of this approach,
\textit{every} function is a candidate for misprediction.
If we were to implement this kind of prediction in a real
application with debloating enabled,
we would need instrumentation and prediction-handling mechanisms
at the entry-point of all functions. This is not realistic from
a performance standpoint.

In summary, this naive approach shows that whole applications seem to
exhibit predictable behavior with even a simple call sequence-based
predictor, but it is also untenable. It shows that by ignoring
the program structure (loops, essentially), the profiling data becomes
unmanageable for normal use cases. There is also no clear
way of handling mispredictions when function calls are simply
flattened into a kind of stream or sequence.

\subsubsection{Program Structure-based Profiling}
Based on our results from the naive scheme, we are 
motivated to look at an approach that takes
the program structure into account.
We adopt a similar static analysis approach as the one in
\cite{decker}, which describes how to identify ``good''
instrumentation points, i.e. points in the program that will
still lead to effective debloating while limiting the impact on performance.
In brief, we can keep overhead low by instrumenting
prediction calls within the application at
loop entry points. These entry points are either at a loop
preheader, or they are at the callsite to a function that is
interprocedurally enclosed by a loop or strongly connected component
through recursion. We call the set of functions that is
statically reachable from one of these entry points as a
``sub-callgraph'' (\textbf{SCG}). Note that indirect
function calls must also be instrumented.

The trained model answers the following question:
\textit{Given a particular sub-callgraph and features,
which functions of the sub-callgraph are expected to
execute dynamically at that program point?}
More precisely, the target of the model is an integer ID
that represents a subset of functions in the sub-callgraph
that will execute. We call this the ``predicted sub-callgraph'' (\textbf{PSCG}).
The features of the model are the arguments to the function entry-point
of the SCG, or, when a loop is the entry-point to an SCG,
the arguments to the parent function of that loop.

The purpose of our profiling instrumentation is to provide these
features and target values as sample data for training
(recall the profiling stage in
Figure \ref{figure:artd-high-level-flow-diagram}).
To capture the features, we instrument a logging function at the
entry-point to SCGs that prints the arguments to the function
being called (or, in the case of loops, the arguments to
its parent function). To capture the target values, we
trace all functions -- as we would in a naive, sequence-based
scheme, but with the critical difference that we also
provide call stack information in the trace.
The call stack information allows us to reconstruct
the functions that are executed within the SCG, and therefore
to assign IDs (the target values) for training.

As for the compiler instrumentation pass for the release stage,
it is much simpler than this. Once the hard work for the profiling
stage is done, the release instrumentation simply needs to
insert a model-invoke function at the entry-point to every SCG.

\subsubsection{Model}
We use decision trees as our model for several reasons.
Empirical evidence in \cite{blankit} already shows that they can be
effective predictors of
execution path behavior based on dynamic program features (such
as the runtime arguments passed to a function call).
Second, decision trees can be embedded as if-else branches.
This is important for performance, because the new framework must
support model invocation within loops and critical paths
of the program. In contrast, 
invoking a complex model in a third-party library at runtime to
predict the upcoming functions would be prohibitively expensive.

Note that the training data for \shortname is also fundamentally different
from the input or test cases used in other debloating frameworks
that are either unsound or specialize the application.
That is, the training data for \shortname is not a stand-in for
the desired features, functionality, or configuration, nor must
its fidelity be perfect because of potential soundness issues.
For example, in Razor \cite{razor}, the test cases are used to uncover
the features and ultimately yield a useful binary that hopefully will
not crash. Similarly, in Chisel \cite{chisel}, the hope is that the
training stage fully captures the desired application/user features;
if it does not, then the Chisel model will make unsound cuts to the program.
In contrast, the training data for \shortname, though it ought to be good,
does not affect the soundness of the resulting binary. The worst-case
scenario for \shortname is that it always mispredicts, but it would
still be sound because of its rectification component.

\subsection{Rectification} \label{sec:rectification}

The previous section describes how the model and its
supporting instrumentation are designed to activate
functions at runtime of predicted sub-callgraphs (PSCGs)
rather than entire sub-callgraphs (SCGs) in order to
further reduce the program's attack surface.
Rectification is about how to handle mispredictions so that the program does
not crash. That is, when the program
attempts to execute code outside of the activated PSCG,
the rectification component must activate the necessary code in the SCG.

Rectifying mispredictions at runtime has two characteristics:
\begin{enumerate}
    \item It guarantees soundness of the application when using
    imperfect predictions to guide which code to map executable.
    \item It adds security beyond that of on-the-fly, binary-level
    debloating techniques like BlankIt \cite{blankit} that effectively
    have to check for rectification at \textit{all} function calls.
\end{enumerate}

Regarding (1), rectification guarantees soundness by ensuring that before a
function executes, it is guaranteed to have an active mapping.
When that function is part of the PSCG (i.e. the predicted set),
this is trivially true (no rectification
is needed). When that function is not part of the PSCG,
it implies that execution is flowing across a boundary between
the PSCG and the SCG; and because all PSCGs are
\textit{known, static targets of the trained model},
\shortname can statically guarantee that all such edges will
be instrumented with rectification instrumentation.
Thus, the \shortname runtime will catch and map those functions
as active before continuing execution.
Regarding (2), rectification yields additional
security beyond schemes like BlankIt because \shortname instruments
only at select points in the callgraph, which
we call ``rectification points'' (\textbf{RPs}).
In schemes such as BlankIt, all function calls are allowed to
proceed, though mispredictions trigger a mitigation strategy.
In \shortname, however, \textit{only function calls at RPs are
allowed to proceed} (and mispredictions will trigger a
path-checking strategy, detailed in the next subsection).
We show an example first to help illustrate these properties and
aid understanding of our rectification algorithm.

\begin{figure}[ht]
    \centering
    \includegraphics[scale=0.2]{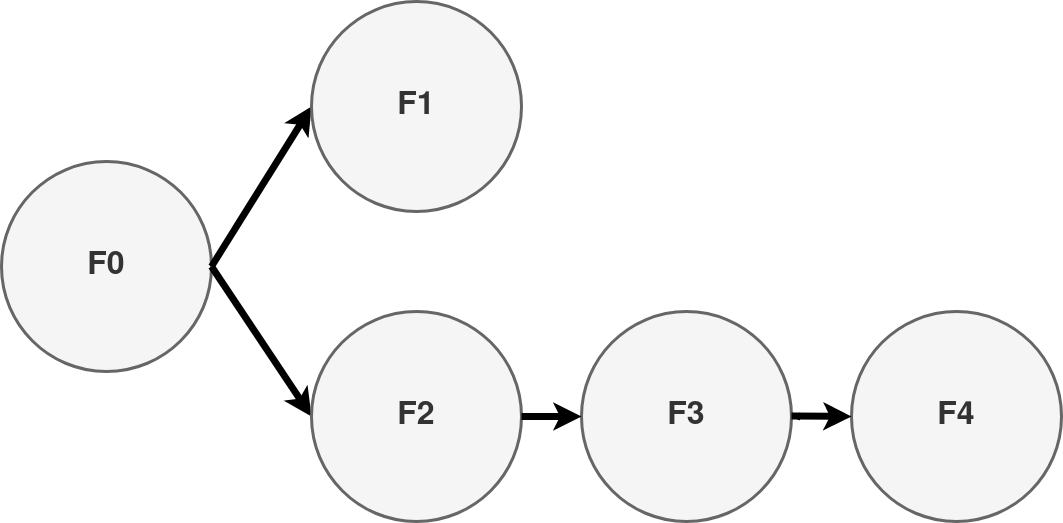}
    \caption{Example callgraph for demonstrating rectification.}
    \label{figure:rectification-example-1}
\end{figure}

Figure \ref{figure:rectification-example-1} shows an example callgraph of five functions,
$F0-F4$. For the purposes of this example, we assume that this callgraph forms an
SCG and that the root of the SCG is $F0$. Assume also that our training
leads to two PSCGs that get called within the SCG:
$A=\{F0, F1, F2\}$ and $B=\{F0, F2, F3\}$.
We will show that in this example, we need 3 RPs:
$RP_{A0} = F2 \rightarrow F3$,
$RP_{B0} = F0 \rightarrow F1$, and
$RP_{B1} = F3 \rightarrow F4$.

The high-level reasoning for these RPs is as follows.
The PSCGs will only ever activate set $A$ or set $B$ (never both,
and never anything outside of those sets).
Therefore, whenever the predicted set is $A$, a misprediction means that
$F3$ and $F4$ must be mapped at an RP, namely $RP_{A0}$
(because otherwise a crash could occur).
Similarly, whenever the predicted set is $B$, a misprediction means that
$F1$ and $F4$ must be mapped at an RP; in this case, they are not
connected, so two RPs are required, namely $RP_{B0}$ and $RP_{B1}$.

The exact implementation mechanisms and semantics for RPs can vary.
For example, at $RP_{B0}$, the framework can mark both $F1$
and $F4$ as active (and disable $RP_{B1}$). Alternatively,
it could service only $F1$ and leave $RP_{B1}$ enabled (which
will service $F4$, should it be needed).
Both are reasonable schemes depending
on the expected runtime overheads and prediction accuracy.
This can be refined through further empirical studies, but for the
purposes of this work, we use the former approach.

To return to our prior point that this rectification can provide
security \textit{beyond} that of on-the-fly, binary-level debloating techniques like
BlankIt, notice that all possible prediction sets are known statically.
(Likewise, all misprediction sets are known statically.)
Therefore, if we take the example of prediction set $A$,
we know at build time that a misprediction includes only
functions $F3$ and $F4$, and that $F4$ is dominated by $F3$
in the callgraph. In other words, there is no way for execution
to go from $F0$, $F1$, or $F2$ directly to $F4$, and thus there is no
RP to handle code activation along such a path. 
This ``gate-like'' behavior is not present in techniques like
BlankIt which map any function on demand and depend on a mitigation
strategy to catch any misprediction.


The pseudocode for instrumenting at RPs is presented in Algorithm
\ref{alg:instrument-at-rps}. The algorithm iterates over the PSCGs
and for each one conducts a depth-first search
over its functions. If it finds a callsite to a callee that is not
part of the PSCG, it inserts an RP at the callsite (\texttt{add\_rp}
in Algorithm \ref{alg:instrument-at-rps}).
\texttt{add\_rp} simply instruments a call to the
\shortname runtime, which will actually map the statically known,
remaining functions of the SCG during execution.

\begin{algorithm}[htb]
\caption{Pseudocode for instrumenting at the rectification points for all PSCGs of a given SCG.}
\label{alg:instrument-at-rps}
\begin{algorithmic}[]
\Function{instrument\_at\_rps}{$SCG$, $PSCGs$}
  \For{$PSCG$ in $PSCGs$}
      \State $rectify\_SCG \gets SCG \setminus PSCG$
      \For{$func$ in $PSCG.\Call{dfs\_next\_node}$}
          \For{$callsite$ in $func.\Call{callsites}$}
              \If{$callsite.callee$ not in $PSCG$}
                \State $callsite.\Call{add\_rp}{rectify\_SCG}$
              \EndIf
          \EndFor
      \EndFor
  \EndFor
\EndFunction
\end{algorithmic}
\end{algorithm}

\subsection{Path Checking} \label{sec:path-checking}
\shortname's path-checking component handles
dynamic, ML-based mispredictions based on program properties
derived from static program analysis.
Path checking is a key motivation for this work, because
other prediction mechanisms for debloating (e.g. \cite{chisel,blankit})
do not give any formal guarantees for checking mispredictions.
It offers a security improvement \textit{beyond} that
already provided by the attack surface reduction from the
prediction component, and above the gating benefit from the
rectification component.
The insight is that static analysis can ``package'' a multitude
of useful program properties for the runtime to check
whenever there is a misprediction.
\shortname records a dynamic call trace so that on misprediction, it can
compare that trace against static properties of the program.
This allows \shortname to handle mispredictions with less overhead
and clearer guarantees than a heavy-weight mitigation strategy
like in \cite{blankit}.

The shape of this solution is as follows: At build time, \shortname encodes
static program properties in Datalog, which serves as a deductive
database. Then at runtime, \shortname runs the program with tracing enabled
and on misprediction, queries the database to check that the trace
satisfies the properties.

There are three major challenges to be tackled, which we cover in turn:
\begin{enumerate}
    \item What static properties are useful, and how do we define them?
    \item How do we capture those properties in a practical compiler setting?
    \item How do we perform the checking online with minimal interference?
\end{enumerate}

\subsubsection{Defining Static Program Properties}

For path validation, we propose to capture a static invariant called
the \textbf{ensue relation}, which we can use for basic validation
on mispredictions. We define the ensue relation as follows:
\begin{definition}
    A call sequence is an ordered list of callsite IDs,
    as they are encountered in an executing program.
\end{definition}
\begin{definition}
    An ensue relation is a pair of adjacent callsite IDs
    within a valid call sequence.
\end{definition}

An example will make this clearer. Consider the code in
Listing \ref{code:path-check-ex1}. Possible call sequences include:
$0.main \rightarrow 1.A \rightarrow 5.C \rightarrow 3.D \rightarrow 4.E$
and
$0.main \rightarrow 2.B \rightarrow 3.D \rightarrow 4.E$.
Ensue relations include:
$ensue(0.main, 1.A)$, 
$ensue(0.main, 2.B)$, 
$ensue(1.A, 5.C)$, 
$ensue(5.C, 3.D)$, 
$ensue(2.B, 3.D)$, 
$ensue(3.D, 4.E)$.
Notice that the ensue relation is determined by both the callgraph and
the control flow graph.


\begin{lstlisting}[
  xleftmargin=.05\textwidth,
  xrightmargin=.05\textwidth,
  caption={Example code for demonstrating various static relations.},
  label=code:path-check-ex1,
  language=C
]
main() {
  if (...)
    1. A()
  else
    2. B()
  3. D()
  4. E()
}
A(){ 5. C() } B(){}C(){}D(){}E(){}
\end{lstlisting}

We identify seven static program properties
in order to fully capture the ensue relation:

\textbf{head(f, i)}:
The first subroutine invoked by function $f$ is at callsite $i$.
Referring to Listing \ref{code:path-check-ex1} as an example,
the head relations are:
  $head(main, 1)$,
  $head(main, 2)$, and
  $head(A, 5)$.

\textbf{tail(f, i)}:
The last subroutine invoked by function $f$ is at callsite $i$.
The tail relations are:
  $tail(main, 4)$,
  $tail(A, 5)$.

\textbf{next(f, i, j)}:
The callsites $i$ and $j$ are adjacent in function $f$.
The next relations are:
  $next(main, 1, 3)$,
  $next(main, 2, 3)$, and
  $next(main, 3, 4)$.

\textbf{leaf(f)}:
The function $f$ does not invoke any subroutine.
The leaf relations are:
  $leaf(B)$, $leaf(C)$, $leaf(D)$, and $leaf(E)$.

\textbf{belong(i, f)}:
The function called at callsite $i$ is function $f$.
The belong relations are:
  $belong(0, main)$,
  $belong(1, A)$,
  $belong(2, B)$,
  $belong(3, D)$,
  $belong(4, E)$, and
  $belong(5, C)$.

\textbf{last(f, i)}:
The function called at callsite $i$ is the last function to be
pushed to the stack before the stack starts popping all the
way down to function $f$. This is a derived relation:
    \begin{align*}
    & last(f, i) \dlimpl tail(f, i), belong(i, g), leaf(g). \\
    & last(f, i) \dlimpl tail(f, j), belong(j, g), last(g, i).
    \end{align*}
   
\textbf{ensue(i, j)}:
The function called at callsite $i$ immediately precedes the function called
at callsite $j$ in some valid call sequence.
This is a derived relation:
    \begin{align*}
    & ensue(i, j) \dlimpl head(f, j), belong(i, f). \\
    & ensue(i, j) \dlimpl next(g, i, j), belong(i, f), leaf(f). \\
    & ensue(i, j) \dlimpl next(g, k, j), belong(k, f), last(f, i).
    \end{align*}

Next we show how to compute the ensue() relation by composing the above. 

\subsubsection{Capturing Static Program Properties}

We implement the first five of these properties within LLVM.
The remaining two ($last$ and $ensue$) are derived relations that we
capture solely in Datalog once we prime its database with the other
five facts.

The $leaf$ and $belong$ relations are straightforward. \shortname simply needs
an LLVM pass that scans all functions and enumerates their callsites. Each
callsite $i$ in function $f$ is added as $belong(i, f)$, and any function $f$
with no callsites is added as $leaf(f)$.

The $head$ and $tail$ relations could be written as classic graph traversals
of the CFG and reverse CFG.
The $next$ relation, however, is non-trivial once the CFG is
moderately complex. Thus, for these three relations, we lean on dataflow techniques
and define $IN$ and $OUT$ transfer functions.

We define two sets of transfers functions, $AHEAD$
and $BEHIND$.
These transfer functions capture the callsites
immediately ahead and behind of a given point in the CFG, which is precisely
the information needed to capture the $head$, $tail$, and $next$ relations.
Intuitively, to find out the callsites immediately ahead of a block,
the information in front must be propagated backward.
Similarly, to find out the callsites immediately behind
a block, the information behind must be propagated forward.
Thus, the $AHEAD$ transfer functions (backward flow) are:
\begin{align*}
  & OUT\_AHEAD[B] = \cup_{S\ succ\ of\ B}\ IN\_AHEAD[S] \\
  & IN\_AHEAD[B]  = Callsite_B\ ||\ OUT\_AHEAD[B]
\end{align*}

\noindent The $BEHIND$ transfer functions (forward flow) are:
\begin{align*}
  & IN\_BEHIND[B]  = \cup_{P\ pred\ of\ B}\ OUT\_BEHIND[P] \\
  & OUT\_BEHIND[B] = Callsite_B\ ||\ IN\_BEHIND[B]
\end{align*}

We do not show the pseudocode for building these sets, but they are similar
to algorithms for other dataflow equations.
The algorithm iterates over the basic blocks of a function,
performing updates to the sets according to the above. Iteration ceases once
there are no more changes to the sets.

After the \shortname release pass creates these sets, it must actually build
the $head$, $tail$, and $next$ facts from them. 
The pseudocode for this is shown in Algorithm \ref{alg:build-dl-facts}.
Briefly, the $head$ of a function is simply $IN\_AHEAD$ of the entry block.
$tail$ and $next$ require us to traverse the CFG and visit each block.
The local functions \texttt{t\_visit} and \texttt{n\_visit} build the
$tail$ and $next$ facts, respectively.
$tail$ consists of all callsites in the $OUT\_BEHIND$ set of a function's exit blocks.
A $next$ relation is added between every block's callsite and the callsites in
its $OUT\_AHEAD$ set.

\begin{algorithm}[htb]
\caption{Capturing $head$, $tail$, and $next$ sets from the transfer functions.}
\label{alg:build-dl-facts}
\begin{algorithmic}[]
\Function{build\_dl\_facts}{$F$}
  \State $head[F] \gets IN\_AHEAD[F.\Call{getEntryBlock}{ }]$
  \State $stack.\Call{push}{ F.entry\_block }$
  \While{$!stack.\Call{empty}$}
    \State $B \gets stack.\Call{pop}$
    \If{$B.visited$}
        \algorithmiccontinue
    \EndIf
    \State $\Call{t\_visit}{B}$
    \State $\Call{n\_visit}{B}$
    \State $B.visited \gets true$
    \State $stack.\Call{push}{ B.successors }$
  \EndWhile
  \\
  \Function{t\_visit}{$B$}      
    \If{$B.\Call{isExitBlock}$}
      \State $tail[F].\Call{insert}{OUT\_BEHIND[B]}$
    \EndIf
  \EndFunction
  \\
  \Function{n\_visit}{$B$}      
    %
    %
    %
    \If{$B.callsite$}
      \For{$callsite$ in $OUT\_AHEAD[B]$}
        \State $next[F].\Call{insert}{B.callsite, callsite}$
      \EndFor
    \EndIf
  \EndFunction
\EndFunction
\end{algorithmic}
\end{algorithm}

\subsubsection{Online Path Checking}
To perform online path checking,
we leverage the program properties defined and computed above.
We also need tracing support at function granularity.
There are multiple options for this, including
hardware support (e.g. Intel PT),
dynamic binary instrumentation (e.g. Intel Pin \cite{Pin}),
or simply compiler instrumentation.
For the \shortname prototype, we use the compiler to add basic
tracing. Then during execution, and on misprediction,
the \shortname runtime must verify that
the (traced) call sequence history is a valid chain of ensue relations.
The checking can be done sequentially (i.e. in the critical
path of the application), in parallel, or written to file
and processed offline. In our experiments, we assume the checks
are done sequentially.
The goal is to check that all adjacent pairs
of executed functions in the call sequence history adhere to
the ensue relation. \shortname maintains the ensue relations in
memory as a map of sets (i.e. to check $ensue(a,b)$, one can
check if $b$ is in the set returned by the map $ensue$ at key $a$).
Under no-attack scenarios, all queries should return true.


\subsection{Putting It All Together}
Now that we have described the framework's technical components in detail,
we want to step back to clarify again how they fit together.
From the user's standpoint, they build an application with \shortname
and profile it using representative inputs. This creates a log
that gets automatically fed into a learner for training a model.
The user then recompiles their application with \shortname,
which embeds the model into the program.
This final binary contains the necessary instrumentation
and linking so that when it runs, \shortname will dynamically reduce
its attack surface.

To achieve this attack surface reduction, several of \shortname's
features work in concert. As the program executes,
the function calls are tracked. Then at key points
in the callgraph, the prediction mechanism is invoked,
and \shortname enables the upcoming region of code that it expects
to execute (i.e. the predicted sub-callgraph, PSCG). As mentioned,
the output of the predictor is a single-value integer ID,
which represents the \textit{set} of functions in the PSCG.
By design, this set is always a subset of the SCG. In other
words, at runtime \shortname is periodically enabling upcoming
functions that it expects will execute, and these functions
are always within the statically known and reachable portion
from the current point in the callgraph.

When execution flows outside of the current PSCG, this is
a misprediction, and it must be handled. \shortname checks
that its call sequence history
conforms to the database of ensue relations.
That is, if a pair of calls is an invalid ensue relation,
then \shortname has detected an attack (or bug) and can act accordingly
(alert, exit, etc.). If, on the other hand, the call sequence
history is verified against the ensue relations, then the
program continues as normal.
\section{Evaluation}\label{sec:evaluation}

All experiments are run on one machine using an
AMD Ryzen 7 with 32GB RAM.
All experiments are on SPEC CPU 2017.
We use LLVM v11, python v3.6.9 with sklearn v0.22.1,
and Datalog v2.6.
Our evaluation is focused on the following questions:
\begin{enumerate}
    \item Does a predictive technique on whole applications
    significantly improve attack surface reduction in terms of
    gadget counts? 
    \item What is the performance overhead of \shortname?
    \item What are the rectification and prediction characteristics,
    both offline and online (e.g. frequency and accuracy)?
    \item Does capturing the ensue relation with Datalog scale for
    complex program behavior?
\end{enumerate}

\subsection{Gadget Reduction}
The gadget reductions are shown in Table \ref{table:spec2017-gadget-reduction}.
This is for all gadget types (ROP, JOP, COP, and special-purpose) reported
by the GSA tool \cite{notsofast}.
Because of \shortname's dynamic behavior, the gadget reductions fluctuate
over the duration of the program, depending on the active PSCGs and SCGs.
Thus, the minimum and maximum columns represent the worst- and best-case
reductions during each application's execution.
The lowest reduction across all benchmarks at any point is for \texttt{mcf}
with 24\% reduction; and the highest reduction is nearly 100\% (after
rounding) for both \texttt{perlbench} and \texttt{xalancbmk}.
On average, however, the minimum and maximum reductions tend to fall between
60 and 90\%, with the average being 82.5\%.

Compared with other state-of-the-art techniques, \shortname's gadget
reductions appear to be an advancement. Of the other sound debloating
techniques, Piece-wise debloats 72.9\% of total gadgets from libraries;
BlankIt debloats 97.8\% ROP gadgets from libraries; LMCAS debloats 55.9\%
ROP gadgets from applications; and Decker debloats 73.2\% total gadgets
from applications.
Thus, \shortname is nearly a 10\% improvement over the nearest, closely
related prior work, Decker, and more than 25\% improvement over the other
sound, whole application-debloating technique, LMCAS.

On closer inspection, \shortname appears to be acting (as intended) to restrict
the functions within the SCGs to a narrow predicted subset. For example,
on average for SPEC CPU 2017, functions and interprocedurally outermost loop
headers statically reach $\sim$45 and $\sim$35 other functions, respectively.
Though these averages collapse the differences
among the benchmarks, it is useful to compare these with the average
cardinality of the PSCGs reported by \shortname, which is $\sim$6 functions;
similarly the complement sets of the PSCGs hold $\sim$35 functions on average.
In other words, \shortname's trained model is much more restrictive than
enabling the otherwise large, statically reachable function sets such as in Decker;
and the benefit to attack  surface reduction is limited by the
mispredictions and unseen inputs not available during training.

\begin{table}[ht!]
 \centering
  \caption{SPEC CPU 2017 total gadget reduction as a percentage (higher is better).}
 \begin{tabular}{|l|l|l|l|}
  \hline
  \textbf{Application} & \bf Min & \bf Max &  \bf Avg \\
  \hline
  \hline
    perlbench	&	38.8	&	100.0	&	67.9 \\
    gcc	&	45.1	&	99.8	&	79.9 \\
    mcf	&	24.8	&	70.0	&	59.5 \\
    namd	&	77.0	&	96.1	&	92.2 \\
    parest	&	90.4	&	99.9	&	98.3 \\
    povray	&	59.1	&	99.5	&	78.5 \\
    lbm	&	53.0	&	65.5	&	59.2 \\
    omnetpp	&	73.1	&	99.3	&	90.3 \\
    xalancbmk	&	82.8	&	100.0	&	92.9 \\
    x264	&	41.4	&	96.9	&	63.5 \\
    blender	&	93.2	&	99.7	&	99.6 \\
    deepsjeng	&	51.9	&	87.5	&	84.6 \\
    imagick	&	52.6	&	99.4	&	94.8 \\
    leela	&	48.8	&	87.8	&	81.8 \\
    nab	&	79.0	&	99.7	&	94.3 \\
    xz	&	66.4	&	95.3	&	82.6 \\
    AVERAGE	&	61.1	&	93.5	&	82.5 \\
  \hline
  \end{tabular}
  \label{table:spec2017-gadget-reduction}
\end{table}


\subsection{Performance Overhead}

\begin{figure}[ht]
    \centering
    \includegraphics[scale=0.35]{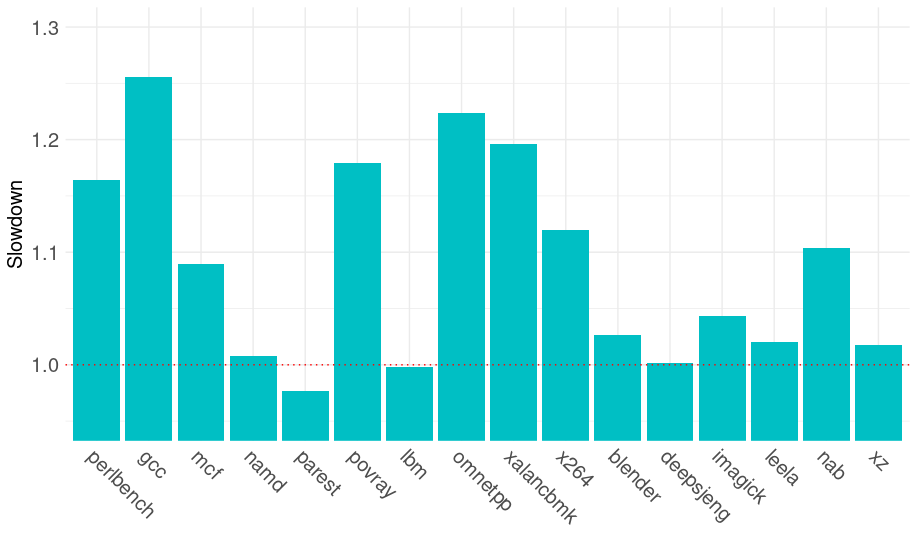}
    \caption{Slowdown for SPEC CPU 2017 using \shortname.}
    \label{figure:spec2017-perf}
\end{figure}

The performance overhead is shown in Figure \ref{figure:spec2017-perf}.
The average overhead is 8.9\%.
The other debloating techniques with runtime components,
Decker and BlankIt, report 5.2\% and 18\% overhead on SPEC, respectively.
(Note that BlankIt protects libraries, however, and that it reports
for SPEC 2006, not 2017). 
\shortname's improved defense over Decker appears to come at a
cost of 3.7\% overhead. Compared with BlankIt, \shortname of
course has less gadget reduction but does not incur the
the heavier slowdowns.

The overheads can be broken down into four groups:
page-mapping instrumentation, prediction and rectification,
tracing support, and ensue querying.
The cost of activating and deactivating pages
(both outside and inside of loops) is similar
to comparable works such as Decker.
In other words, \shortname's additional slowdown
over these techniques is from the other three components.
We isolate and measure their contributions.
Tracing adds $\sim$0.5\%, and there is $\sim$2\%
overhead from the ensue queries. The remaining $\sim$1\% overhead
is due to the prediction and rectification.
We also test with larger ensue
buffers (in the critical path and without blocking), and see
ensue querying add up to $\sim$4\% overhead for 16-element
ensue buffers (vs. a simpler two-element history).
We notice seven of the
SPEC benchmarks have decision trees with max depths over 10
levels: \texttt{perlbench}, \texttt{gcc}, \texttt{parest},
\texttt{omnetpp}, \texttt{xalancbmk}, \texttt{x264},
and \texttt{blender}.
We find this corresponds loosely with worst-case
slowdowns from Figure \ref{figure:spec2017-perf}, which merits further
investigation (i.e. more closely measuring and then checking
how limiting their tree depths affects the overall security and
performance).

Another focus of our performance overhead study is the
contribution from indirect function calls.
Of all the predictions issued by \shortname, $\sim$32\% of these
are due to indirect calls.
More importantly, because they
require instrumentation inside of loops, they are heavy
contributors to the slowdown.
\shortname ``caches'' active code pages inside of loops
(i.e. keeps them active until loop exit) to mitigate this
overhead. Thus, the prediction cost to performance should
only be paid once per indirect call inside of a loop,
but the instrumentation instructions will still be expensive.
We disable tracing, ensue querying, and indirect call instrumentation
in an additional experiment and see only $\sim$3\% overhead from \shortname
over baseline SPEC CPU 2017. This is partly expected but
also remarkable. It demonstrates the heavy penalty for
instrumenting inside of loops and strongly suggests that
the \shortname profile data could be leveraged to avoid
instrumenting in particularly hot loop sections.

\subsection{Rectification and Prediction}


We capture the percent of callgraph edges with rectification
instrumentation, and we record at runtime how frequently these edges
are traversed. The former metric gives a sense of how many
RPs are actually needed to ensure sound behavior at runtime.
``Lower is better'' does not necessarily hold here.
A high count of RPs could be due to a benchmark's particular callgraph
structure and how it is exercised by its training inputs (which
ultimately determines the predicted sets and thus the RPs);
nevertheless, a low count of RPs implies that there are fewer ``gates''
through which an attack must pass if it is attempting to load
functions outside of the predicted set.

The second metric is a useful way to report the prediction accuracy.
When execution reaches an RP (i.e. when it traverses a callgraph edge
with rectification instrumentation), it is because execution is
flowing outside of the predicted set of functions -- a misprediction.
In the current implementation, the rectification
behavior is to map all remaining functions in that SCG. Therefore,
whenever execution reaches an RP, the misprediction is handled
once (and only once) for that SCG. The misprediction rate is thus
the percentage of all rectification events over all prediction events.

Regarding the former metric, the average percentage of edges that are
instrumented with RPs for SPEC is $\sim$6.3\%.
All but two benchmarks have 10\% or less of their
edges instrumented with RPs. \texttt{lbm} has 13.3\%,
and \texttt{x264} has 44.8\%.
Regarding the second metric,
Table \ref{table:spec-pred-accuracy} reports the frequency of the
prediction events vs. the rectification events. The rightmost column
reports this as a percentage. On average, \shortname must
rectify 3.8\% of all predictions that happen for SPEC CPU 2017.
The worst case is \texttt{leela}, which has only 4 predictions,
1 of which triggers rectification.


\begin{table}[ht!]
 \centering
  \caption{SPEC CPU 2017 prediction and rectification occurences. The rightmost
  column is the percentage of predictions that require rectification.}
 \begin{tabular}{|l|l|l|l|}
  \hline
  \textbf{Application} & \bf \#Predicts & \bf \#Rectifies & \bf \%Rectifies \\
  \hline
  \hline
    perlbench	&	175	&	7	&	4 \\
    gcc	&	449	&	6	&	1.3 \\
    mcf	&	5	&	0	&	0 \\
    namd	&	19	&	0	&	0 \\
    parest	&	777	&	17	&	2.2 \\
    povray	&	43	&	2	&	4.7 \\
    lbm	&	3	&	0	&	0 \\
    omnetpp	&	513	&	1	&	0.2 \\
    xalancbmk	&	342	&	6	&	1.8 \\
    x264	&	60	&	4	&	6.7 \\
    blender	&	8687	&	17	&	0.2 \\
    deepsjeng	&	4	&	0	&	0 \\
    imagick	&	26	&	2	&	7.7 \\
    leela	&	4	&	1	&	25 \\
    nab	&	95	&	2	&	2.1 \\
    xz	&	48	&	2	&	4.2 \\
  \hline
  \end{tabular}
  \label{table:spec-pred-accuracy}
\end{table}

\subsection{Ensue Relations and Datalog}
The \shortname compiler pass outputs a Datalog program which includes
all $head$, $tail$, $next$, $leaf$, and $belong$ facts, along with the derived
relations for $last$ and $ensue$, and, crucially, a single query as its
final line: $ensue(A,B)?$. We execute this Datalog program offline.
Datalog reads all of the facts, derives $last$ and $ensue$, and
returns the result of the final query, which is a list of all ensue
relations.

\begin{table}[ht!]
 \centering
  \caption{Datalog results for SPEC CPU 2017.}
 \begin{tabular}{|l|l|l|l|l|}
  \hline
  \textbf{Application} & \bf Time &  \bf Size & \bf \#Facts &  \bf \#Ensue \\
  \hline
  \hline
    perlbench	&	16m	&	8MB	&	270764	&	392007 \\
    gcc	&	478m	&	232MB	&	1318536	&	11468987 \\
    mcf	&	0m	&	1MB	&	123	&	46 \\
    namd	&	0m	&	1MB	&	1357	&	711 \\
    parest	&	43m	&	4MB	&	152680	&	172228 \\
    povray	&	2m	&	4MB	&	62619	&	175976 \\
    lbm	&	0m	&	1MB	&	73	&	26 \\
    omnetpp	&	2m	&	2MB	&	56165	&	108442 \\
    xalancbmk	&	33m	&	13MB	&	142410	&	667789 \\
    x264	&	0m	&	1MB	&	9145	&	26541 \\
    blender	&	76m	&	11MB	&	227995	&	507865 \\
    deepsjeng	&	0m	&	1MB	&	1336	&	1644 \\
    imagick	&	9m	&	8MB	&	95182	&	414880 \\
    leela	&	0m	&	1MB	&	2164	&	3210 \\
    nab	&	0m	&	1MB	&	1613	&	2419 \\
    xz	&	0m	&	1MB	&	1674	&	1207 \\
  \hline
  \end{tabular}
  \label{table:datalog-results}
\end{table}

The results are shown in Table \ref{table:datalog-results}.
The column headers mean the following:
Time is the length of time for Datalog to produce all ensue
relations for the application, rounded to the nearest minute;
Size is the filesize (reported by \texttt{du -sm}) of the output
from Datalog of all ensue relations; \#Facts is the number
of facts in the Datalog program (input); \#Ensue is the number
of ensue relations Datalog reports (output).

In half of the benchmarks, Datalog completes in less than 1 minute.
The most striking case is \texttt{gcc}, which takes nearly 8 hours to complete.
Though all of this processing is offline, this result shows that
calculating ensue relations for highly complex callgraphs can be
compute-intensive.
Furthermore, the Size result (output filesize) matters for
runtime, because it must be loaded into memory as a fast data structure
and queried during execution.
In all cases but \texttt{gcc}, the data size is reasonable.
The benchmark with the most Datalog facts
after \texttt{gcc} is \texttt{perlbench}.
Though \texttt{gcc} has $\sim$5x more facts than \texttt{perlbench},
its output (in terms of filesize, which is correlated with the number
of ensue relations), is $\sim$30x,
and it takes $\sim$30x longer to run, as well.
In other words, the relative complexity of \texttt{gcc}
is exceptionally high compared with the other benchmarks,
as seen in its ensue count, and this increases its computational
costs for Datalog.

\section{Related Work} \label{sec:related-work}

A number of attack surface reduction techniques
restrict the feature set of the debloated application.
For example, Slimium \cite{slimium} uses a mix of
static and dynamic methods to produce
a reduced version of Chromium.
BinRec \cite{binrec, binrec-workshop} is a binary-level technique
that lifts binaries to LLVM IR and can leverage dynamic traces to
recompile a transformed program;
it supports safety mechanisms to ensure sound execution
outside of the debloated version of the program.
Because feature debloating is typically input-dependent,
the authors in \cite{tradeoffs} introduce metrics and tools
for studying the correctness and robustness of
such methods; they also show how
to improve the generality of debloated programs without too
heavily compromising the code reduction.

There are works in software debloating that are not aimed
specifically at security problems.
They may target
performance \cite{causes_of_bloat,making_sense_debloat,analysis_debloat,four_trends_debloat,finding_utility_debloat,container_debloat,finding_reusable_debloat},
code footprint reduction \cite{Beszedes:2003:SCR:937503.937504,
Debray:2000:CTC:349214.349233,Muth:2001:ALO:370365.370382,
Franz:1997:SB:265563.265576}, and maintenance \cite{debloat-maintenance}.

Lastly, control flow integrity
(CFI) \cite{cfi,xfi} is an orthogonal defense
with many examples in the literature
(e.g. \cite{griffin, uCFI, patharmor, typearmor, pittypat, mcfi, cfi-lb, oscfi}).
CFI attempts to enforce forward and backward control flow transfers to
only legal paths in the CFG and callgraph, which
can thwart code reuse attacks.
Similarly, control pointer integrity (CPI) \cite{cpi,getting-point}
attempts to guarantee the integrity of all code pointers.
Examples in the industry include Microsoft's Control Flow Guard
\cite{windows-cfg} and Return Flow Guard \cite{windows-rfg}
and Intel's Control Flow Enforcement Technology \cite{intel-CET}
with Indirect Branch Tracking \cite{linux-ibt} and a hardware
shadow stack.
There is an equally rich body of research showing these techniques
to be vulnerable \cite{jop-rocket,counterfeit, mcafee-blackhat,bypass-cfg,
bypass-cfg-2, bypass-cfg-3,control-jujutsu,control_flow_bending,cracks},
prompting the need for software debloating research.
\section{Conclusion}\label{sec:conclusion}
This work presents a new framework that combines
machine learning and static analysis to
advance whole-application debloating.
We show how precise predictions can narrow the amount
of active code at runtime, and how static invariants
derived at build time and enforced at runtime can be used
as lightweight checks to filter any true attacks
from the ML model's mispredictions.
The prediction scheme takes advantage of
program structure, and the checking
mechanism is based on valid static call sequences.
The resulting programs that use this framework maintain
full feature support, and all transformations are sound.
We show attacks that escape
Decker but can be caught by the proposed technique. 
We improve attack surface reduction beyond state of the art
(82.5\% total gadget reduction on SPEC CPU 2017)
with reasonable runtime overhead (8.9\%)
and minimal online checking (3.8\% of predictions).

\bibliographystyle{plain}
\bibliography{refs.bib}

\begin{thebibliography}{10}

\bibitem{cfi}
Mart{\'{\i}}n Abadi, Mihai Budiu, {\'{U}}lfar Erlingsson, and Jay Ligatti.
\newblock Control-flow integrity.
\newblock In Vijay Atluri, Catherine~A. Meadows, and Ari Juels, editors, {\em Proceedings of the 12th {ACM} Conference on Computer and Communications Security, {CCS} 2005, Alexandria, VA, USA, November 7-11, 2005}, pages 340--353. {ACM}, 2005.

\bibitem{lmcas}
Mohannad Alhanahnah, Rithik Jain, Vaibhav Rastogi, Somesh Jha, and Thomas~W. Reps.
\newblock Lightweight, multi-stage, compiler-assisted application specialization.
\newblock In {\em 7th {IEEE} European Symposium on Security and Privacy, EuroS{\&}P 2022, Genoa, Italy, June 6-10, 2022}, pages 251--269. {IEEE}, 2022.

\bibitem{binrec}
Anil Altinay, Joseph Nash, Taddeus Kroes, Prabhu Rajasekaran, Dixin Zhou, Adrian Dabrowski, David Gens, Yeoul Na, Stijn Volckaert, Cristiano Giuffrida, Herbert Bos, and Michael Franz.
\newblock Binrec: dynamic binary lifting and recompilation.
\newblock In Angelos Bilas, Kostas Magoutis, Evangelos~P. Markatos, Dejan Kostic, and Margo~I. Seltzer, editors, {\em EuroSys '20: Fifteenth EuroSys Conference 2020, Heraklion, Greece, April 27-30, 2020}, pages 36:1--36:16. {ACM}, 2020.

\bibitem{Beszedes:2003:SCR:937503.937504}
\'{A}rp\'{a}d Besz{\'e}des, Rudolf Ferenc, Tibor Gyim\'{o}thy, Andr{\'e} Dolenc, and Konsta Karsisto.
\newblock Survey of code-size reduction methods.
\newblock {\em ACM Comput. Surv.}, 35(3):223--267, September 2003.

\bibitem{green-bloat}
Suparna Bhattacharya, Kanchi Gopinath, Karthick Rajamani, and Manish Gupta.
\newblock Software bloat and wasted joules: Is modularity a hurdle to green software?
\newblock {\em Computer}, 44(9):97--101, 2011.

\bibitem{jop}
Tyler~K. Bletsch, Xuxian Jiang, Vincent~W. Freeh, and Zhenkai Liang.
\newblock Jump-oriented programming: a new class of code-reuse attack.
\newblock In Bruce S.~N. Cheung, Lucas Chi~Kwong Hui, Ravi~S. Sandhu, and Duncan~S. Wong, editors, {\em Proceedings of the 6th {ACM} Symposium on Information, Computer and Communications Security, {ASIACCS} 2011, Hong Kong, China, March 22-24, 2011}, pages 30--40. {ACM}, 2011.

\bibitem{jop-rocket}
Bramwell Brizendine and Austin Babcock.
\newblock Pre-built jop chains with the jop rocket: Bypassing dep without rop.
\newblock \url{ https://i.blackhat.com/asia-21/Thursday-Handouts/as-21-Brizendine-Babcock-Prebuilt-Jop-Chains-With-The-Jop-Rocket-wp.pdf}, 2021.
\newblock Accessed: 2022 Jun 06.

\bibitem{islessreallymore}
Michael~D. Brown and Santosh Pande.
\newblock Is less really more? towards better metrics for measuring security improvements realized through software debloating.
\newblock In Rob Jansen and Peter A.~H. Peterson, editors, {\em 12th {USENIX} Workshop on Cyber Security Experimentation and Test, {CSET} 2019, Santa Clara, CA, USA, August 12, 2019.} {USENIX} Association, 2019.

\bibitem{notsofast}
Michael~D. Brown, Matthew Pruett, Robert Bigelow, Girish Mururu, and Santosh Pande.
\newblock Not so fast: understanding and mitigating negative impacts of compiler optimizations on code reuse gadget sets.
\newblock {\em Proc. {ACM} Program. Lang.}, 5({OOPSLA}):1--30, 2021.

\bibitem{control_flow_bending}
Nicolas Carlini, Antonio Barresi, Mathias Payer, David Wagner, and Thomas~R. Gross.
\newblock Control-flow bending: On the effectiveness of control-flow integrity.
\newblock In {\em Proceedings of the 24th USENIX Conference on Security Symposium}, SEC'15, pages 161--176, Berkeley, CA, USA, 2015. USENIX Association.

\bibitem{jop2}
Stephen Checkoway, Lucas Davi, Alexandra Dmitrienko, Ahmad{-}Reza Sadeghi, Hovav Shacham, and Marcel Winandy.
\newblock Return-oriented programming without returns.
\newblock In Ehab Al{-}Shaer, Angelos~D. Keromytis, and Vitaly Shmatikov, editors, {\em Proceedings of the 17th {ACM} Conference on Computer and Communications Security, {CCS} 2010, Chicago, Illinois, USA, October 4-8, 2010}, pages 559--572. {ACM}, 2010.

\bibitem{linux-ibt}
Jonathan Corbet.
\newblock Indirect branch tracking for intel cpus.
\newblock \url{https://lwn.net/Articles/889475/}, 2022.
\newblock Accessed: 2022 Oct 21.

\bibitem{Debray:2000:CTC:349214.349233}
Saumya~K. Debray, William Evans, Robert Muth, and Bjorn De~Sutter.
\newblock Compiler techniques for code compaction.
\newblock {\em ACM Trans. Program. Lang. Syst.}, 22(2):378--415, March 2000.

\bibitem{pittypat}
Ren Ding, Chenxiong Qian, Chengyu Song, William Harris, Taesoo Kim, and Wenke Lee.
\newblock Efficient protection of path-sensitive control security.
\newblock In Engin Kirda and Thomas Ristenpart, editors, {\em 26th {USENIX} Security Symposium, {USENIX} Security 2017, Vancouver, BC, Canada, August 16-18, 2017}, pages 131--148. {USENIX} Association, 2017.

\bibitem{debloat-maintenance}
Sebastian Eder, Maximilian Junker, Elmar J{\"{u}}rgens, Benedikt Hauptmann, Rudolf Vaas, and Karl{-}Heinz Prommer.
\newblock How much does unused code matter for maintenance?
\newblock In Martin Glinz, Gail~C. Murphy, and Mauro Pezz{\`{e}}, editors, {\em 34th International Conference on Software Engineering, {ICSE} 2012, June 2-9, 2012, Zurich, Switzerland}, pages 1102--1111. {IEEE} Computer Society, 2012.

\bibitem{xfi}
{\'{U}}lfar Erlingsson, Mart{\'{\i}}n Abadi, Michael Vrable, Mihai Budiu, and George~C. Necula.
\newblock {XFI:} software guards for system address spaces.
\newblock In Brian~N. Bershad and Jeffrey~C. Mogul, editors, {\em 7th Symposium on Operating Systems Design and Implementation {(OSDI} '06), November 6-8, Seattle, WA, {USA}}, pages 75--88. {USENIX} Association, 2006.

\bibitem{control-jujutsu}
Isaac Evans, Fan Long, Ulziibayar Otgonbaatar, Howard Shrobe, Martin Rinard, Hamed Okhravi, and Stelios Sidiroglou-Douskos.
\newblock Control jujutsu: On the weaknesses of fine-grained control flow integrity.
\newblock In {\em Proceedings of the 22Nd ACM SIGSAC Conference on Computer and Communications Security}, CCS '15, pages 901--913, New York, NY, USA, 2015. ACM.

\bibitem{Franz:1997:SB:265563.265576}
Michael Franz and Thomas Kistler.
\newblock Slim binaries.
\newblock {\em Commun. ACM}, 40(12):87--94, December 1997.

\bibitem{griffin}
Xinyang Ge, Weidong Cui, and Trent Jaeger.
\newblock {GRIFFIN:} guarding control flows using intel processor trace.
\newblock In Yunji Chen, Olivier Temam, and John Carter, editors, {\em Proceedings of the Twenty-Second International Conference on Architectural Support for Programming Languages and Operating Systems, {ASPLOS} 2017, Xi'an, China, April 8-12, 2017}, pages 585--598. {ACM}, 2017.

\bibitem{chisel}
Kihong Heo, Woosuk Lee, Pardis Pashakhanloo, and Mayur Naik.
\newblock Effective program debloating via reinforcement learning.
\newblock In {\em Proceedings of the 2018 ACM SIGSAC Conference on Computer and Communications Security}, CCS '18, pages 380--394, New York, NY, USA, 2018. ACM.

\bibitem{debloat-art}
Curt Hibbs, Steve Jewett, and Mike Sullivan.
\newblock {\em The Art of Lean Software Development}.
\newblock O'Reilly, 2009.

\bibitem{lstm}
Sepp Hochreiter and Jürgen Schmidhuber.
\newblock Long short-term memory.
\newblock {\em Neural computation}, 9:1735--80, 12 1997.

\bibitem{uCFI}
Hong Hu, Chenxiong Qian, Carter Yagemann, Simon Pak~Ho Chung, William~R. Harris, Taesoo Kim, and Wenke Lee.
\newblock Enforcing unique code target property for control-flow integrity.
\newblock In {\em Proceedings of the 2018 ACM SIGSAC Conference on Computer and Communications Security}, CCS '18, pages 1470--1486, New York, NY, USA, 2018. ACM.

\bibitem{oscfi}
Mustakimur Khandaker, Wenqing Liu, Abu Naser, Zhi Wang, and Jie Yang.
\newblock Origin-sensitive control flow integrity.
\newblock In Nadia Heninger and Patrick Traynor, editors, {\em 28th {USENIX} Security Symposium, {USENIX} Security 2019, Santa Clara, CA, USA, August 14-16, 2019}, pages 195--211. {USENIX} Association, 2019.

\bibitem{cfi-lb}
Mustakimur Khandaker, Abu Naser, Wenqing Liu, Zhi Wang, Yajin Zhou, and Yueqiang Cheng.
\newblock Adaptive call-site sensitive control flow integrity.
\newblock In {\em {IEEE} European Symposium on Security and Privacy, EuroS{\&}P 2019, Stockholm, Sweden, June 17-19, 2019}, pages 95--110. {IEEE}, 2019.

\bibitem{binrec-workshop}
Taddeus Kroes, Anil Altinay, Joseph Nash, Yeoul Na, Stijn Volckaert, Herbert Bos, Michael Franz, and Cristiano Giuffrida.
\newblock Binrec: Attack surface reduction through dynamic binary recovery.
\newblock In {\em Proceedings of the 2018 Workshop on Forming an Ecosystem Around Software Transformation}, FEAST '18, page 8–13, New York, NY, USA, 2018. Association for Computing Machinery.

\bibitem{cpi}
Volodymyr Kuznetsov, L\'{a}szl\'{o} Szekeres, Mathias Payer, George Candea, R.~Sekar, and Dawn Song.
\newblock Code-pointer integrity.
\newblock In {\em Proceedings of the 11th USENIX Conference on Operating Systems Design and Implementation}, OSDI'14, pages 147--163, Berkeley, CA, USA, 2014. USENIX Association.

\bibitem{getting-point}
Volodymyr Kuznetsov, L{\'a}szl{\'o} Szekeres, Mathias Payer, George Candea, and Dawn Song.
\newblock Poster : Getting the point (er) : On the feasibility of attacks on code-pointer integrity.
\newblock 2015.

\bibitem{cracks}
Yuan Li, Mingzhe Wang, Chao Zhang, Xingman Chen, Songtao Yang, and Ying Liu.
\newblock Finding cracks in shields: On the security of control flow integrity mechanisms.
\newblock In Jay Ligatti, Xinming Ou, Jonathan Katz, and Giovanni Vigna, editors, {\em {CCS} '20: 2020 {ACM} {SIGSAC} Conference on Computer and Communications Security, Virtual Event, USA, November 9-13, 2020}, pages 1821--1835. {ACM}, 2020.

\bibitem{windows-cfg}
Microsoft.
\newblock Control flow guard for platform security.
\newblock \url{https://learn.microsoft.com/en-us/windows/win32/secbp/control-flow-guard}, 2022.
\newblock Accessed: 2022 Oct 21.

\bibitem{making_sense_debloat}
Nick Mitchell, Edith Schonberg, and Gary Sevitsky.
\newblock Making sense of large heaps.
\newblock In Sophia Drossopoulou, editor, {\em {ECOOP} 2009 - Object-Oriented Programming, 23rd European Conference, Genoa, Italy, July 6-10, 2009. Proceedings}, volume 5653 of {\em Lecture Notes in Computer Science}, pages 77--97. Springer, 2009.

\bibitem{four_trends_debloat}
Nick Mitchell, Edith Schonberg, and Gary Sevitsky.
\newblock Four trends leading to java runtime bloat.
\newblock {\em {IEEE} Software}, 27(1):56--63, 2010.

\bibitem{causes_of_bloat}
Nick Mitchell and Gary Sevitsky.
\newblock The causes of bloat, the limits of health.
\newblock In Richard~P. Gabriel, David~F. Bacon, Cristina~Videira Lopes, and Guy L.~Steele Jr., editors, {\em Proceedings of the 22nd Annual {ACM} {SIGPLAN} Conference on Object-Oriented Programming, Systems, Languages, and Applications, {OOPSLA} 2007, October 21-25, 2007, Montreal, Quebec, Canada}, pages 245--260. {ACM}, 2007.

\bibitem{Muth:2001:ALO:370365.370382}
Robert Muth, Saumya~K. Debray, Scott Watterson, and Koen De~Bosschere.
\newblock Alto: A link-time optimizer for the compaq alpha.
\newblock {\em Softw. Pract. Exper.}, 31(1):67--101, January 2001.

\bibitem{Nergal}
Nergal.
\newblock The advanced return-into-lib(c) exploits: Pax case study. phrack magazine.
\newblock \url{http://phrack.org/issues/58/4.html}, 2001.
\newblock Accessed: 2021 Oct 10.

\bibitem{Nethercote:2007:VFH:1250734.1250746}
Nicholas Nethercote and Julian Seward.
\newblock Valgrind: A framework for heavyweight dynamic binary instrumentation.
\newblock In {\em Proceedings of the 28th ACM SIGPLAN Conference on Programming Language Design and Implementation}, PLDI '07, pages 89--100, New York, NY, USA, 2007. ACM.

\bibitem{mcfi}
Ben Niu and Gang Tan.
\newblock Modular control-flow integrity.
\newblock In Michael F.~P. O'Boyle and Keshav Pingali, editors, {\em {ACM} {SIGPLAN} Conference on Programming Language Design and Implementation, {PLDI} '14, Edinburgh, United Kingdom - June 09 - 11, 2014}, pages 577--587. {ACM}, 2014.

\bibitem{Pin}
H.~Patil, R.~Cohn, M.~Charney, R.~Kapoor, A.~Sun, and A.~Karunanidhi.
\newblock Pinpointing representative portions of large intel \#174; itanium \#174; programs with dynamic instrumentation.
\newblock In {\em Microarchitecture, 2004. MICRO-37 2004. 37th International Symposium on}, pages 81--92, Dec 2004.

\bibitem{decker}
Chris Porter, Sharjeel Khan, and Santosh Pande.
\newblock Decker: Attack surface reduction via on-demand code mapping.
\newblock In Tor~M. Aamodt, Natalie D.~Enright Jerger, and Michael~M. Swift, editors, {\em Proceedings of the 28th {ACM} International Conference on Architectural Support for Programming Languages and Operating Systems, Volume 2, {ASPLOS} 2023, Vancouver, BC, Canada, March 25-29, 2023}, pages 192--206. {ACM}, 2023.

\bibitem{blankit}
Chris Porter, Girish Mururu, Prithayan Barua, and Santosh Pande.
\newblock Blankit library debloating: getting what you want instead of cutting what you don't.
\newblock In Alastair~F. Donaldson and Emina Torlak, editors, {\em Proceedings of the 41st {ACM} {SIGPLAN} International Conference on Programming Language Design and Implementation, {PLDI} 2020, London, UK, June 15-20, 2020}, pages 164--180. {ACM}, 2020.

\bibitem{razor}
Chenxiong Qian, Hong Hu, Mansour Alharthi, Simon Pak~Ho Chung, Taesoo Kim, and Wenke Lee.
\newblock {RAZOR:} {A} framework for post-deployment software debloating.
\newblock In Nadia Heninger and Patrick Traynor, editors, {\em 28th {USENIX} Security Symposium, {USENIX} Security 2019, Santa Clara, CA, USA, August 14-16, 2019}, pages 1733--1750. {USENIX} Association, 2019.

\bibitem{slimium}
Chenxiong Qian, Hyungjoon Koo, ChangSeok Oh, Taesoo Kim, and Wenke Lee.
\newblock Slimium: Debloating the chromium browser with feature subsetting.
\newblock In Jay Ligatti, Xinming Ou, Jonathan Katz, and Giovanni Vigna, editors, {\em {CCS} '20: 2020 {ACM} {SIGSAC} Conference on Computer and Communications Security, Virtual Event, USA, November 9-13, 2020}, pages 461--476. {ACM}, 2020.

\bibitem{debloat-study}
Anh Quach, Rukayat Erinfolami, David Demicco, and Aravind Prakash.
\newblock A multi-os cross-layer study of bloating in user programs, kernel and managed execution environments.
\newblock In Taesoo Kim, Cliff Wang, and Dinghao Wu, editors, {\em Proceedings of the 2017 Workshop on Forming an Ecosystem Around Software Transformation, FEAST@CCS 2017, Dallas, TX, USA, November 3, 2017}, pages 65--70. {ACM}, 2017.

\bibitem{piece-wise}
Anh Quach, Aravind Prakash, and Lok Yan.
\newblock Debloating software through piece-wise compilation and loading.
\newblock In {\em 27th {USENIX} Security Symposium ({USENIX} Security 18)}, pages 869--886, Baltimore, MD, 2018. {USENIX} Association.

\bibitem{ropgadget}
Ropgadget v5.4.
\newblock \url{https://github.com/JonathanSalwan/ROPgadget}, 2018.
\newblock Accessed: 2021 Oct 10.

\bibitem{pcop}
AliAkbar Sadeghi, Salman Niksefat, and Maryam Rostamipour.
\newblock Pure-call oriented programming {(PCOP):} chaining the gadgets using call instructions.
\newblock {\em J. Comput. Virol. Hacking Tech.}, 14(2):139--156, 2018.

\bibitem{bypass-cfg-2}
Morten Schenk.
\newblock Bypassing control flow guard in windows 10.
\newblock \url https://improsec.com/tech-blog/bypassing-control-flow-guard-in-windows-10.

\bibitem{ropper}
Sascha Schirra.
\newblock Ropper.
\newblock \url{https://github.com/sashs/ropper}, 2021.
\newblock Accessed: 2021 Sept 10.

\bibitem{counterfeit}
Felix Schuster, Thomas Tendyck, Christopher Liebchen, Lucas Davi, Ahmad{-}Reza Sadeghi, and Thorsten Holz.
\newblock Counterfeit object-oriented programming: On the difficulty of preventing code reuse attacks in {C++} applications.
\newblock In {\em 2015 {IEEE} Symposium on Security and Privacy, {SP} 2015, San Jose, CA, USA, May 17-21, 2015}, pages 745--762. {IEEE} Computer Society, 2015.

\bibitem{DBLP:conf/usenix/SewardN05}
Julian Seward and Nicholas Nethercote.
\newblock Using valgrind to detect undefined value errors with bit-precision.
\newblock In {\em Proceedings of the 2005 {USENIX} Annual Technical Conference, April 10-15, 2005, Anaheim, CA, {USA}}, pages 17--30. {USENIX}, 2005.

\bibitem{rop}
Hovav Shacham.
\newblock The geometry of innocent flesh on the bone: return-into-libc without function calls (on the x86).
\newblock In Peng Ning, Sabrina De~Capitani di~Vimercati, and Paul~F. Syverson, editors, {\em Proceedings of the 2007 {ACM} Conference on Computer and Communications Security, {CCS} 2007, Alexandria, Virginia, USA, October 28-31, 2007}, pages 552--561. {ACM}, 2007.

\bibitem{intel-CET}
Vedvyas Shanbhogue, Deepak Gupta, and Ravi Sahita.
\newblock Security analysis of processor instruction set architecture for enforcing control-flow integrity.
\newblock In {\em Proceedings of the 8th International Workshop on Hardware and Architectural Support for Security and Privacy}, HASP '19, New York, NY, USA, 2019. Association for Computing Machinery.

\bibitem{trimmer}
Hashim Sharif, Muhammad Abubakar, Ashish Gehani, and Fareed Zaffar.
\newblock {TRIMMER:} application specialization for code debloating.
\newblock In Marianne Huchard, Christian K{\"{a}}stner, and Gordon Fraser, editors, {\em Proceedings of the 33rd {ACM/IEEE} International Conference on Automated Software Engineering, {ASE} 2018, Montpellier, France, September 3-7, 2018}, pages 329--339. {ACM}, 2018.

\bibitem{mcafee-blackhat}
Bing Sun, Jin Liu, and Chong Xu.
\newblock How to survive the hardware-assisted control-flow integrity enforcement.
\newblock \url{ https://i.blackhat.com/asia-19/Thu-March-28/bh-asia-Sun-How-to-Survive-the-Hardware-Assisted-Control-Flow-Integrity-Enforcement.pdf}, 2019.
\newblock Accessed: 2022 Jun 06.

\bibitem{bypass-cfg-3}
Sam Thomas.
\newblock Object oriented exploitation: New techniques in windows mitigation bypass.
\newblock \url https://www.slideshare.net/\_s\_n\_t/object-oriented-exploitation-new-techniques-in-windows-mitigation-bypass.

\bibitem{expressiveness_retlibc}
Minh Tran, Mark Etheridge, Tyler Bletsch, Xuxian Jiang, Vincent Freeh, and Peng Ning.
\newblock On the expressiveness of return-into-libc attacks.
\newblock In Robin Sommer, Davide Balzarotti, and Gregor Maier, editors, {\em Recent Advances in Intrusion Detection}, pages 121--141, Berlin, Heidelberg, 2011. Springer Berlin Heidelberg.

\bibitem{patharmor}
Victor van~der Veen, Dennis Andriesse, Enes G{\"{o}}ktas, Ben Gras, Lionel Sambuc, Asia Slowinska, Herbert Bos, and Cristiano Giuffrida.
\newblock Practical context-sensitive {CFI}.
\newblock In Indrajit Ray, Ninghui Li, and Christopher Kruegel, editors, {\em Proceedings of the 22nd {ACM} {SIGSAC} Conference on Computer and Communications Security, Denver, CO, USA, October 12-16, 2015}, pages 927--940. {ACM}, 2015.

\bibitem{typearmor}
Victor van~der Veen, Enes G{\"{o}}ktas, Moritz Contag, Andre Pawoloski, Xi~Chen, Sanjay Rawat, Herbert Bos, Thorsten Holz, Elias Athanasopoulos, and Cristiano Giuffrida.
\newblock A tough call: Mitigating advanced code-reuse attacks at the binary level.
\newblock In {\em {IEEE} Symposium on Security and Privacy, {SP} 2016, San Jose, CA, USA, May 22-26, 2016}, pages 934--953. {IEEE} Computer Society, 2016.

\bibitem{windows-rfg}
Danny Wei, Lywang, and FlowerCode.
\newblock Return flow guard.
\newblock \url{https://xlab.tencent.com/en/2016/11/02/return-flow-guard/}, 2016.
\newblock Accessed: 2022 Oct 21.

\bibitem{tradeoffs}
Qi~Xin, Qirun Zhang, and Alessandro Orso.
\newblock Studying and understanding the tradeoffs between generality and reduction in software debloating.
\newblock In {\em 37th {IEEE/ACM} International Conference on Automated Software Engineering, {ASE} 2022, Rochester, MI, USA, October 10-14, 2022}, pages 99:1--99:13. {ACM}, 2022.

\bibitem{container_debloat}
Guoqing Xu and Atanas Rountev.
\newblock Detecting inefficiently-used containers to avoid bloat.
\newblock In {\em Proceedings of the 31st ACM SIGPLAN Conference on Programming Language Design and Implementation}, PLDI '10, pages 160--173, New York, NY, USA, 2010. ACM.

\bibitem{finding_reusable_debloat}
Guoqing~(Harry) Xu.
\newblock Finding reusable data structures.
\newblock In Gary~T. Leavens and Matthew~B. Dwyer, editors, {\em Proceedings of the 27th Annual {ACM} {SIGPLAN} Conference on Object-Oriented Programming, Systems, Languages, and Applications, {OOPSLA} 2012, part of {SPLASH} 2012, Tucson, AZ, USA, October 21-25, 2012}, pages 1017--1034. {ACM}, 2012.

\bibitem{finding_utility_debloat}
Guoqing~(Harry) Xu, Nick Mitchell, Matthew Arnold, Atanas Rountev, Edith Schonberg, and Gary Sevitsky.
\newblock Finding low-utility data structures.
\newblock In Benjamin~G. Zorn and Alexander Aiken, editors, {\em Proceedings of the 2010 {ACM} {SIGPLAN} Conference on Programming Language Design and Implementation, {PLDI} 2010, Toronto, Ontario, Canada, June 5-10, 2010}, pages 174--186. {ACM}, 2010.

\bibitem{analysis_debloat}
Guoqing~(Harry) Xu, Nick Mitchell, Matthew Arnold, Atanas Rountev, and Gary Sevitsky.
\newblock Software bloat analysis: finding, removing, and preventing performance problems in modern large-scale object-oriented applications.
\newblock In Gruia{-}Catalin Roman and Kevin~J. Sullivan, editors, {\em Proceedings of the Workshop on Future of Software Engineering Research, FoSER 2010, at the 18th {ACM} {SIGSOFT} International Symposium on Foundations of Software Engineering, 2010, Santa Fe, NM, USA, November 7-11, 2010}, pages 421--426. {ACM}, 2010.

\bibitem{bypass-cfg}
Zhang Yunha.
\newblock Bypass control flow guard comprehensively.
\newblock \url https://www.blackhat.com/docs/us-15/materials/us-15-Zhang-Bypass-Control-Flow-Guard-Comprehensively-wp.pdf.

\end{thebibliography}

\end{document}